\documentclass[twocolumn,aps,prx,amsmath,amssymb,superscriptaddress]{revtex4}

\usepackage{amssymb,color,ulem}
\usepackage{latexsym}
\usepackage{graphicx}
\usepackage{amsmath}
\usepackage{setspace}
\usepackage{fancyhdr}
\usepackage{graphics}
\usepackage{lastpage}
\usepackage{graphics}
\usepackage{comment}

\begin{document}

\title{Incoherent chiral-induced spin selectivity}

\author{Lior Oppenheim}
\affiliation{Racah Institute of Physics, Hebrew University, Jerusalem 91904, Israel}
\affiliation{Department of Condensed Matter Physics, Weizmann Institute of Science, Rehovot 76100, Israel}
\author{Karen Michaeli}
\affiliation{Department of Condensed Matter Physics, Weizmann Institute of Science, Rehovot 76100, Israel}


\begin{abstract}
The observation of spin-dependent transport through organic chiral structures has sparked numerous fundamental and applicative questions ranging from biology to spintronics.  By now, there is a broad consensus that the effect results from the combination of spin-orbit coupling and the systems' unique geometry. However, two key challenges remain. Firstly, accounting for the magnitude of the measured effect in light of the weak spin-orbit coupling in organic systems. Secondly, understanding its observation not only in tunneling-dominated short molecules, but also in longer ones where phonon-assisted hopping via localized states is operative. We focus on the latter and find that localization and the complete loss of coherence due to phonons do not impede strong spin polarization of charge carriers passing through the molecule. Moreover, hopping decouples the energy scale for observing spin-selective transport from the magnitude of the spin-orbit coupling. Thus, our result may explain the observation of large spin selectivity at room temperature and under large applied voltages. 

\end{abstract}

\maketitle

\subsection*{Introduction}

Structures with a well-defined chirality (handedness) arise in nature on all length scales. Examples range from spiral galaxies to crystals and small molecules that cannot be mapped onto their mirror image by any continuous transformation. In solids, broken inversion symmetry may promote the atomic spin orbit coupling (SOC) to a macroscopic property. Phases of matter such as helimagnets~\cite{Bak}, topological insulators~\cite{Kane} and Weyl semimetals~\cite{Yan} all rely on electronic energy bands with substantial SOC. In biomolecules, chirality is prevalent. It characterizes, e.g., amino acids, sugars, and enzymes, all of which occur in living organisms almost exclusively with a single handedness. While many biological and pharmaceutical implications of this homochirality are understood, its effect on electronic properties is only beginning to emerge.



Over the past two decades, experiments consistently observed strongly spin-dependent transport in chiral molecules without any magnetic components ~\cite{review1,review2}. This prominent effect, named chiral-induced spin-selectivity (CISS), has been probed in a variety of different setups. The most direct ones are scattering~\cite{Ray,Goehler,Abendroth,Kiran,Zacharias} and magnetoresistance measurements~\cite{Xie,Kettner,Diez-Perez,Fontanesi}. In the former, electrons are photo-excited from a metallic substrate and pass through a layer of chiral molecules before their current intensity and spin-polarization is measured. This measurement directly probes the spin-dependent transmission probabilities through the molecules. A spin polarization of more than 60\% has been observed in this setup~\cite{Goehler}. In the latter, the electric resistance is measured for molecules connecting two electrodes, one of which is magnetized. The currents observed for opposite signs of the magnetization along the central axis of the molecule are dramatically different. For example, the voltages at which they cross the noise threshold may differ by as much as one Volt~\cite{Xie}.

\begin{figure}[b]
       \includegraphics[width=0.45\textwidth]{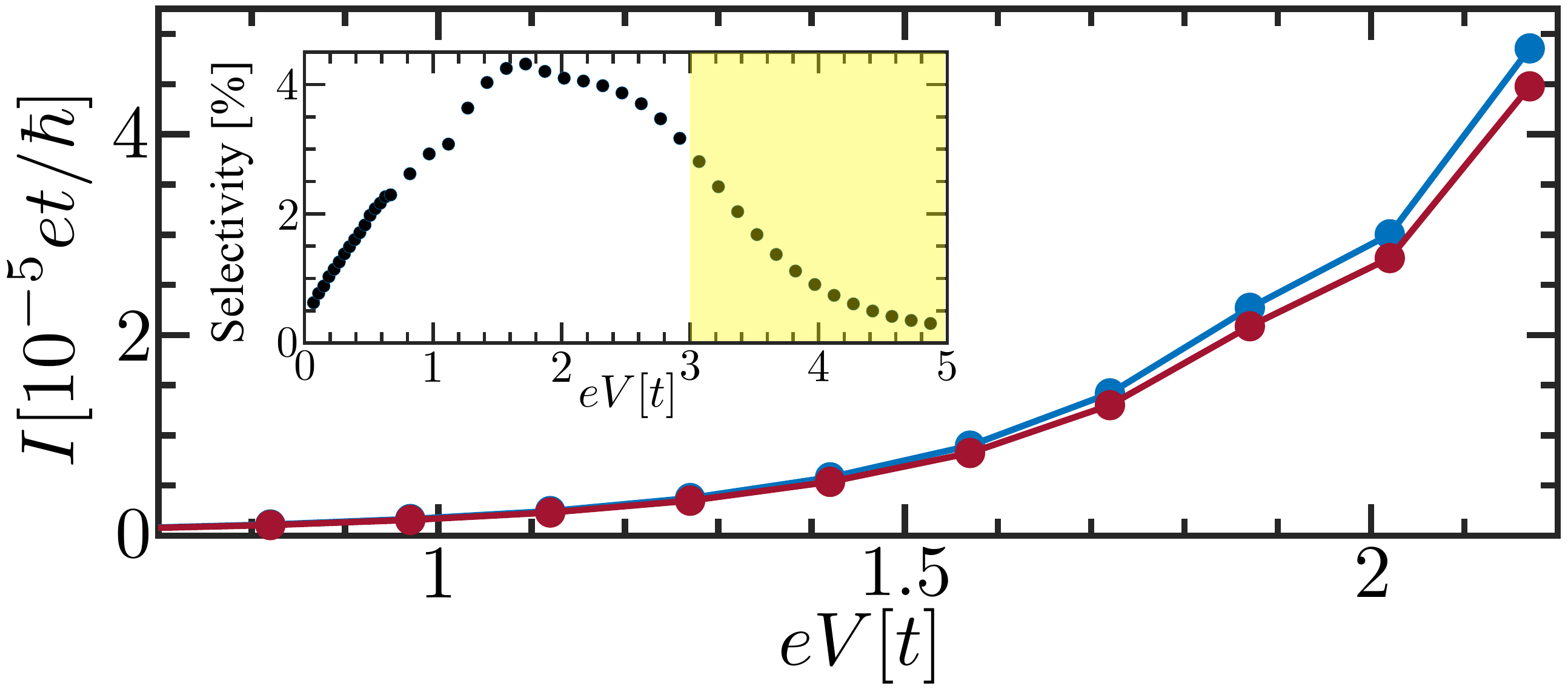}
                \caption{\textbf{Current-voltage characteristic of a moleucle connected to one magnetic and one normal lead}.  The magnet has a moment $+M$ (\textcolor{blue}{blue}) or $-M$ (\textcolor{red}{red}) along the molecule axis. At zero voltage, the chemical potential of the leads lies deep in the band gap of the molecule.  The inset shows the selectivity, $[I(+M)-I(-M)]/[I(+M)+I(-M)]$, which initially increases with voltage, but drops once the chemical potential enters the conduction band (indicated in yellow). 
                 }\label{IV}
\end{figure}

\begin{figure*}
       \includegraphics[width=1\textwidth]{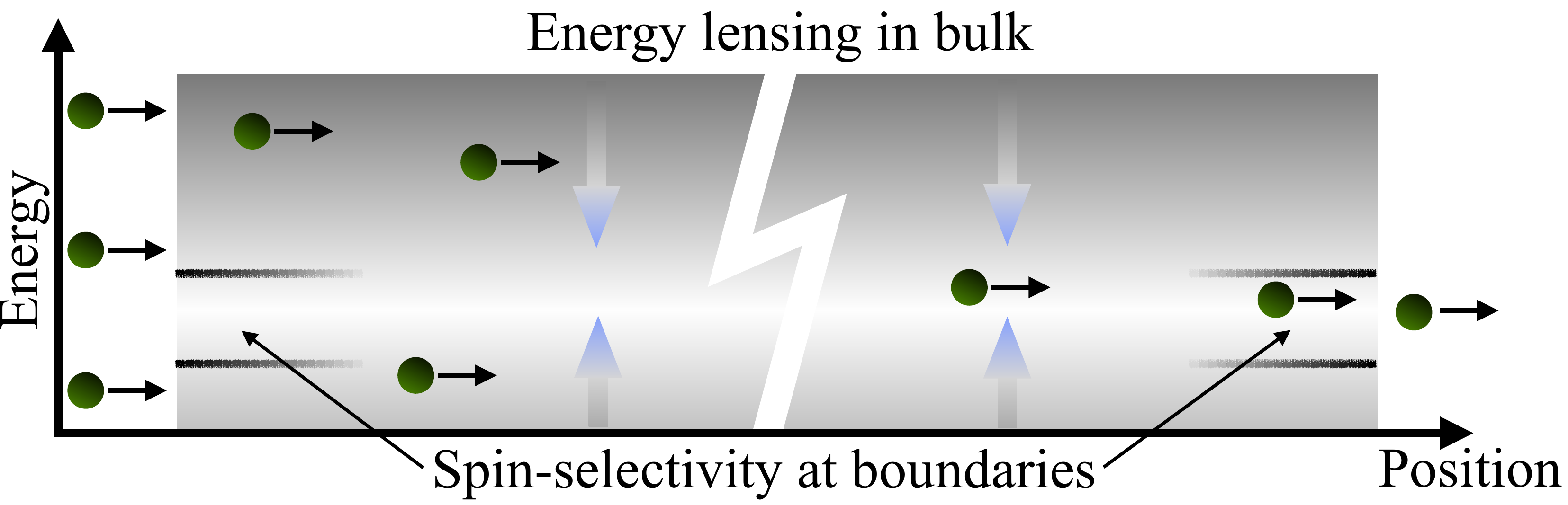}
\caption{\textbf{Mechanism of spin-polarization in the phonon-assisted hopping regime.} The entry and exit of electrons to the molecule is spin selective in a narrow energy window. Phonon-assisted hopping through the bulk of the molecule focuses the electron energies toward the transport energy. Strong spin polarization arises when this energy is close to the spin-selective window, which is typically the case. 
}\label{fig1}
\end{figure*}

 The basic phenomenology of CISS indicates that an electron carrying a given spin has unequal probabilities of passing from left to right and from right to left. By contrast, time-reversal symmetry (TRS) requires that probabilities of electrons of opposite spins to be transmitted in opposite directions be equal in the absence of magnetic fields. Formally, the symmetry between states of different helicity $\textit{h}\equiv\hat{v}_{s}\cdot \vec s=\pm1/2$, i.e., the projection of the electron spin, $\vec s$ on its velocity, $\vec{v}_{s}$,  is broken. Theories suggested that this asymmetry stems from a combination of the curved geometry and SOC~\cite{Yeganeh,Medina,Gutierrez,Guo,Eremko,Rai,Guo2,Medina2,Matityahu,Avishai,Michaeli,Per,Ora}. On the single-particle level, the magnitude of the asymmetry is limited by the strength of the SOC, which is on the order of millielectron Volts for organic molecules. By contrast, CISS is observed at room temperature and under strong applied voltages.

Initial experiments mainly focused on short organic chiral molecules (up to several nanometers) such as double-stranded DNA~\cite{Goehler}, oligopeptides~\cite{Kettner}, and helicenes~\cite{Kiran,Zacharias}.  Recently, long-range spin-dependent transport (up to micrometers) has been observed in superhelical polymer microfibers~\cite{Yan2020}, multiheme electron conduits~\cite{El-Naggar2019} and two-dimensional chiral hybrid organic-inorganic perovskites~\cite{Vardeny2019,Sun2020,Beard2020}. These systems all contain chiral organic molecules through which charge must flow for the system to conduct. Their magnetoresistance reflects a spin selectivity comparable to shorter molecules and suggests a common origin of CISS in both cases. By contrast,  transport through short or long organic structures occurs via very different mechanisms. In the former, charges propagates via coherent tunneling. In the latter, phonon assisted hopping through localized states is typically the dominant effect~\cite{NitzanBook,Berlin,Ref13,Ref14}.

Existing theoretical works studied coherent tunneling through regular molecular chains (atoms distributed equidistantly on a helix) with extended electronic states. On the single-particle level, spin-selectivity occurs only within an energy window set by the SOC. Interaction effects may enhance this window and thus account for the difference in energy scales~\cite{Diaz,Fransson,Qu,Fransson2}. We here focus on the fate of CISS when electronic states are strongly localized. Specifically, we ask under what conditions CISS  survives inhomogeneities, and analyze the role of phonon-assisted hopping for spin-dependent transport.

We find that the asymmetry between helicities leads to weaker back-scattering~\cite{Brouwer} and a longer localization length than in achiral systems. Moreover up to relatively strong disorder, the exponential tails of the localized states are mostly comprised of a single helicity. Coherent transmission is dominated by these tails and thus, substantial spin-dependent charge transfer persists. More specifically, the spin polarization is moderately reduced, but the associated energy window is increased at the same time. 

At the measurement temperatures, transport through long molecules with localized states is dominated by phonon-assisted processes. This regime was analyzed in Ref.~\cite{Diaz2} for a molecular bridge with broken TRS. There, phonon-assisted hopping was found to enhance the spin polarization compared to direct tunneling. We study here the hopping regime in non-magnetic chiral molecules. Due to TRS, hopping within the bulk cannot induce any spin polarization. In addition, since spin is not conserved due to SOC, any memory of a particle's spin is lost after a single hop. We find that charge carriers nevertheless become strongly spin-polarized upon passing through a chiral molecule. Moreover, this polarization is almost independent of the particles initial energy, unlike the tunneling case.

An essential part of our theoretical analysis of CISS effect in the hopping regime was to develop a framework for studying a molecule bridging between two electrodes. This method allows us to connect between our theoretical work and the standard experimental method for observing CISS---magnetoresistance measurements of a chiral molecule coupled to at least one magnetic lead. The resulting current-voltage characteristic shown in Fig.~\ref{IV}, which resembles the experimental data~\cite{Xie}, is one of the main achievements of our work.

We show that the strong spin polarization in the hopping regime arises from the combination of bulk and boundary effects. Spin selectivity occurs at the entry (and exit) of particles to (from) the molecule. Its underlying mechanism is the same as in tunneling, and it is likewise restricted to a limited energy window. Phonon-assisted hopping in the bulk focuses the particle's energy towards the \textit{transport energy}~\cite{Koehler}. The polarization of the exiting particles is thus solely determined by this energy, which is typically within the window of significant spin selectivity (See Fig.~\ref{fig1} for an illustration). This transport mechanism decouples the energy scale for observing CISS from the strength of the SOC. It can thus explain the observation of spin polarization upon scattering through long molecules at room temperature. We demonstrate by an explicit calculation of the magnetoresistance, that phonon-assisted hopping also gives rise to current-voltage characteristics (Fig.~\ref{IV}) consistent with the experimental data.




\subsection*{Model}

Our study is based on a molecular bridge connecting two electrodes (or a donor and an acceptor). The same framework is also widefly used to study charge-transfer through generic molecules without non-trivial spin physics~\cite{Ref4,Ref5}. Spin selectivity arises when SOC is introduced into a helically-shaped bridge.  These two ingredients are central to all theoretical model of CISS. Explicitly, we use the Hamiltonian
\begin{align}\label{mol}
H_{\text{mol}}&=H_0 + H_{\text{SOC}}+H_\text{phonons}.
\end{align}
Here, we focus on non-magnetic molecules with only nearest neighbor coupling, i.e.,
\begin{align}\label{tightbinding}
H_0=&\sum_{n=1}^{N}
\sum_{m,s }\varepsilon_{n,m}c_{n,m,s}^{\dag}c_{n,m,s}
\\ &-\sum_{ n=1 }^{N-1}
\sum_{ m,m',s}
t_{n,m,m'}c_{n+1,m,s}^{\dag}c_{n,m',s}\hspace{-0.6mm}+\text{H.c}.\nonumber
\end{align}
Here, $c_{n,m,s}^{\dag}$ creates a spin $s=\uparrow,\downarrow$ electron on site $n$ of a bridge with length $L=Na_0$, where $a_0$ is the intersite distance. The quantum number $m$ encodes additional local degrees of freedom such as atomic orbitals~\cite{Ora} or chain-number in a double-stranded helix~\cite{Guo}. Following Refs.~\cite{Michaeli,Per,Ora}, the low energy Hamiltonian consists of only two $p$-orbitals, $m=\pm1$, with quantization axis tangential to the helix. Furthermore, on-site energies are equal, $\varepsilon_{n,m}=\varepsilon_{n}$, and mixing is negligible, $t_{n,m,m'}=t_{n}\delta_{m,m'}$.

The nature of the electronic states on a molecular bridge, like the one modeled by Eq.~\ref{tightbinding}, strongly depends on various local parameters. For example, in the absence of SOC and electron-phonon interactions,  uniform on-site energies $\varepsilon_{n}=\varepsilon$ and inter-site couplings $t_n=t$ give rise to  bands of extended states with energies between $\varepsilon-2t$ to $\varepsilon+2t$. Consequently, an incoming electron with energy $E$ in this window is transmitted with a probability of order unity.

The effective SOC for a helix  is  
\begin{align}\label{SOC}
H_{\text{SOC}}=&\Delta_{\text{SOC}}
\sum_{n=1}^{N}
\sum_{m,s,s'}mc_{n,m,s}^{\dag}\bigg[-\frac{b}{4\pi{R}}\sigma_{s,s'}^{z}\\\nonumber
&+ie^{- i \chi 2\pi{n}/\tilde{R}}\sigma_{s,s'}^{+}+c.c.\bigg]c_{n,m,s'},
\end{align}
where $\vec \sigma$ are Pauli matrices, $\tilde R/a_0$ is the number of sites in each turn of the helix, and $\chi = \pm$ denotes the handedness of the molecule. The pitch $b$ and radius of the helix $R$ satisfy $\tilde{R}=\sqrt{(2\pi R)^2+b^2}$. $\Delta_{\text{SOC}}$ is proportional to the atomic SOC (on the order of $5-10$meV). Crucially, the appearance of the time-reversal odd quantum number $m$ as a prefactor ensures that $H_{\text{SOC}}$ is time-reversal symmetric and the molecule remains non-magnetic. Eq.~\ref{SOC} has a similar origin as the coupling between isospin and coordinate in bent graphene~\cite{Geim}. Unless specified otherwise, we take $\tilde{R}=4a_0$ and $\Delta_{\text{SOC}}=0.1t$; the latter corresponds to SOC of about $10$meV in a band of width $\sim400$meV, in agreement with Ref.~\cite{Binghai}. We drop the term proportional to $\sigma_z$ in the SOC, since it has very little quantitative effect on spin transport as long as the pitch and the radius of the helix are comparable in length. If they differ significantly, the molecule looses its helical form and we do not expect to see CISS in any case.

Finally, the phonons and their coupling to the electrons are described by the standard Fr\"ohlich Hamiltonian~\cite{Bryksin}
\begin{align}\label{Frohlich}
	H_\text{phonons} =&\sum_{q,\alpha}\omega_{q,\alpha}b^{\dagger}_{q}b_{q}\\
	&+\sum_{k,q,m,s}\gamma_{k,q}c^{\dagger}_{k,m,s}c_{k-q,m,s}b_{q}+\text{H.c}.
\end{align}
Here $k,q$ are lattice momenta and $b_q^{\dag}$ creates a phonon.

Before addressing the role of imhomogeneities and phonons, we briefly describe CISS on an ideal bridge with $\varepsilon_{n}=\varepsilon$, $t_n=t$. The eigenstates of $H_0$ alone are extended for all $\sigma,m$ at energies between $\varepsilon-2t$ and $\varepsilon+2t$. Consequently, an incoming electron with energy $E$ in this window is transmitted with a probability of order unity. Including $H_\text{SOC}$ opens gaps for half of the electronic states in two windows, one in the lower and one in the upper half band. The remaining states have primarily one helicity---the spins carried by electrons with opposite velocities along the helix axis are almost perfectly anti-aligned.  Thus, there is a strong asymmetry between helicities in the partial gaps, and consequently, the transmission probability of an electron at energy $\epsilon$ within these windows  is strongly spin-dependent.

The helicity of the states on the lower half band is opposite to that in the upper half band, and they are determined by the handedness of the molecule. Therefore, spin-dependent transport only occurs when charges tranfer preferentially via one half of the band, as is usually the case.  Crucially, the helicity in the partial gap is independent of the orbital $m=\pm1$, and their contributions to the spin-selective transmission \textit{add}.  
For a detailed derivation and  analysis of the model, see Refs.~\cite{Michaeli,Per,Ora}. This type of model captures the  CISS effect qualitatively. Yet, the partial gaps are set by the  atomic SOC, which results in a narrow energy window compared to the temperature and voltages relevant for the experiments.

\subsection*{Localization in chiral structures}

Unlike the idealized case described above, organic molecules typically exhibit strong variations in $\varepsilon_n$  and $t_n$. These inhomogeneities arise from the chemical composition of the molecule and the surrounding environment.   In addition to static structural variations,  there are thermal fluctuations of the nuclear degrees of freedom~\cite{Ref11}, i.e., phonons. Generically there are both slow phonons with frequencies much lower than the inverse transfer time of electrons through the molecule as well as fast ones. The former have a similar effect as static inhomogeneities on the transmission of individual electrons~\cite{Ref12}. In the case of DC currents, they amount to a built-in averaging over thermal disorder. Faster phonons play a fundamentally different role, which we analyze in detail below. All (quasi-) static inhomogeneities mix electronic states with different energies and velocities. In particular, backscattering between counter-propagating states results in exponentially localized electronic states with an energy-dependent localization length $\ell_\text{loc}(E)$. At zero temperature, electron transmission occurs through direct tunneling and its probability decays exponentially with distance at all energies.

\begin{figure}
       \includegraphics[width=0.5\textwidth]{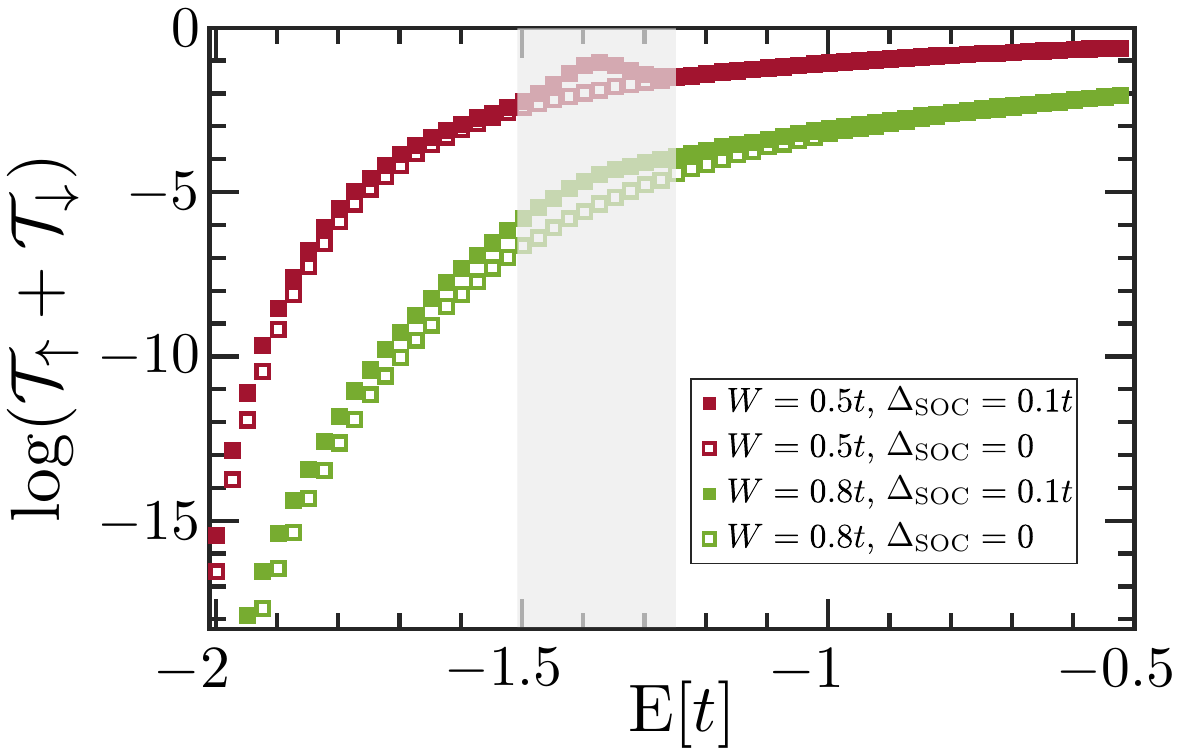}
                \caption{\textbf{Total transmission through inhomogeneous chiral molecules}.  We show the disorder-averaged logarithm of the total transmission $T[E]$ for electrons tunneling through a molecule of length $L=150a_0$ with two disorder strengths in the presence and absence of SOC.  The energies where $\Delta_\text{SOC}$ opens a partial-gap  in the homogeneous limit are indicated in gray. For relatively weak disorder, localization is suppressed in the vicinity of this energy window. For stronger disorder, SOC increases the localization length over a larger energy range.  }\label{fig2}
\end{figure}

We now compute the spin-dependent transmission probability  for coherent tunneling through the localized states. In the model of Eq.~\ref{mol} we incorporate all (quasi-) static inhomogeneities  via $\varepsilon_n=\varepsilon+U_n$ with $U_n$ drawn from a uniform distribution of  uncorrelated site energies in the domain $[-W,W]$. We verified that disorder in the nearest-neighbor coupling $t_n$ and orbital mixing have similar effects as $U_n$.  We use the transfer matrix method~\cite{Soukoulis} to determine the probability $T^\text{coh}_{\sigma,\sigma'}(E)$ of an incoming electron with $E,\sigma'$ to exit with spin $\sigma$. For both weakly localized ($W\ll t$) and strongly localized ($W\approx t$) states, we find that the total transmission $\sum_{\sigma,\sigma'}T^\text{coh}_{\sigma,\sigma'}$ is substantially enhanced by the SOC term, see Fig.~\ref{fig2}. The weaker localization may be attributed to a suppression of backscattering within the partial gap.

\begin{figure}
       \includegraphics[width=0.5\textwidth]{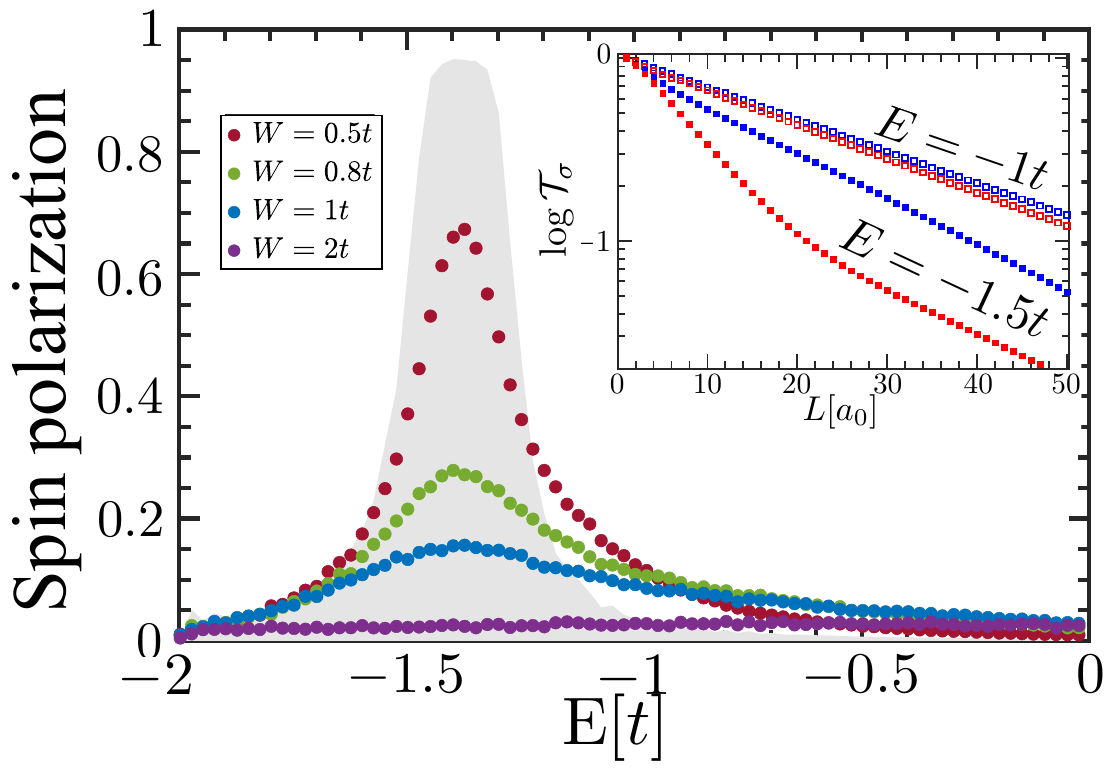}
                \caption{\textbf{CISS in inhomogeneous chiral molecules}. We present the spin polarization ${\cal P}^\text{coh}(E)$ for tunneling through molecules of length $L=150a_0$ with various disorder strengths; the clean-limit polarization is shown in gray. Up to relatively strong disorder, significant spin polarization persists. While ${\cal P}^\text{coh}(E)$ decreases near the peak, it increases at the tails. The inset shows the transmissions ${\cal T}^\text{coh}_{\uparrow}$ (\textcolor{blue}{blue}) and  ${\cal T}^\text{coh}_{\downarrow}$ (\textcolor{red}{red}). The two interchange upon reversing the direction of transmission. The ${\cal T}^\text{coh}_{\downarrow}$ data reflects two length scales, $\lambda\propto \Delta_\text{SOC}^{-1}$, and  $\ell_\text{loc}(E)$, while ${\cal T}^\text{coh}_{\uparrow}$ is set by the latter only. Consequently, the tails of the localized states are dominated by one helicity for $\lambda \ll \ell_\text{loc}(E)$.  }\label{fig3}
\end{figure}

In Fig.~\ref{fig3} we show the spin polarization of the \textit{outgoing} electrons
\begin{align}
{\cal P}^\text{coh}(E)&\equiv\frac{{\cal T}^\text{coh}_{\uparrow}(E)-{\cal T}_{\downarrow}^\text{coh}(E)}{{\cal T}^\text{coh}_{\uparrow}(E)+{\cal T}^\text{coh}_{\downarrow}(E)},
\end{align}
where ${\cal T}^\text{coh}_{\sigma_\text{out}}(E)=\sum_{\sigma_\text{in}}T_{\sigma_\text{in},\sigma_\text{out}}^\text{coh}(E)$. The polarization decreases with disorder, but remains substantial up to $W \approx t$. For molecules much longer than $\ell_\text{loc}$, the energy window where spin-polarization occurs broadens in comparison to the clean limit, $W\rightarrow0$.
To understand the survival of spin-polarization in the presence of disorder, we show the length dependence of ${\cal T}^\text{coh}_{\sigma}(E)$ in Fig.~\ref{fig3}. For energies within the partial gap ${\cal T}^\text{coh}_{\downarrow}$ decays much faster than ${\cal T}^\text{coh}_{\uparrow}$. The two are interchanged for transmission in the opposite directions. Thus, the exponential tails of the localized states are mostly comprised of a single helicity. Some imbalance between the helicities remains even at energies far from the partial gap and thus accounts for the disorder-broadened window of spin polarization.
The inset of Fig.~\ref{fig3} clearly shows two length scales for ${\cal T}^\text{coh}_{\downarrow}$. The initial strong decay is determined by the inverse partial gap, i.e., $\lambda =a_0(t/\Delta_{\text{SOC}} )\sin[\pi a_0/\tilde R]$. The weaker decay at longer distances is due to disorder. The polarization becomes negligible when these two length scales are comparable.


In summary, coherent tunneling through chiral molecules can exhibit significant spin-dependence up to strong disorder. While inhomogeneities that are unavoidable in organic molecules suppress the achievable polarization, they also  extend the energy window where it is observed. Thus, disorder alone can partially explain the discrepancy between the magnitude of SOC and the temperature/voltages where CISS is observed. However, coherent tunneling dominates transport through long molecules only at extremely low temperatures~\cite{NitzanBook}. At higher temperatures, charge transfer occurs primarily via sequence of phonon-assisted hopping events between localized states~\cite{NitzanBook,Berlin,Ref13,Ref14}.


\subsection*{Phonon-assisted hopping on a molecular bridge}

The coupling between electrons and phonons allows to transfer energy and momentum between them and is described by the Fr\"{o}hlich Hamiltonian Eq.~\eqref{Frohlich}. The same model is widely used to examine the role of phonons in various contexts, such as strong and weak polarons. However, it cannot be solved exactly and thus requires approximations that are appropriate for the limit of interest. Here, our focus lies on transport through highly disordered molecules at temperatures where phonon-assisted hopping dominates. In such systems, all states are well localized and electrons can conduct by thermal activation \cite{
NitzanBook,Berlin,Ref4,Ref5}. Specifically, in the Variable Range Hopping (VRH) regime, electrons propagate between localized states by absorbing or emitting  phonons; in between each events, the electrons lose coherence. Miller and Abrahams \cite{miller_1960} derived the effective hopping rate from state $i$ with energy $E_i$ to state $j$ with energy $E_j$ to be
\begin{equation}
\label{MillerAbrahams}
	\nu_{i,j}=\nu_{0}\alpha_{i,j}\begin{cases} e^{-\beta\left(E_j-E_i\right)} & E_j>E_i\\1 & \text{else} \end{cases}.
\end{equation}
Here, $\beta$ is the inverse temperature and the typical attempt rate $\nu_0$ is set by the electron-phonon interaction and the phonon spectrum. $\alpha_{i,j} =\int dx\left|\psi_{i}(x)\right|^2\left|\psi_{j}(x)\right|^2$ is the spatial overlap between the electronic densities at the initial and final states. These equilibrium probabilities directly determine bulk transport properties such as particle diffusivity and conductivity. 


We now examine the implications of VHR in the setup of the scattering experiments. There, a current-carrying steady-state is achieved by injecting an electron beam into one side of the molecule and collecting it at the other side. The current intensity and the spin polarization of the outgoing beam are then measured. Scattering experiments are customarily described by attaching the system~\cite{Datta} to infinite leads, which incorporate the extended nature of the incoming and final states. Thus, we attach $H_\text{mol}$ [Eq.~\eqref{mol}] to two leads, i.e., we study 
\begin{align}
	H_\text{scatter}=&H_{\text{L}}+H_{\text{mol}}+H_{\text{R}}\\
	+&\gamma \sum_{m,s}\left[c_{1,m,s}^{\dag}d_{L,1,m,s}\hspace{-0.6mm}+c_{N,m,s}^{\dag}d_{R,1,m,s}\hspace{-0.6mm}+\text{H.c}.\right].\nonumber
	\label{total}
\end{align}
The left and right leads are governed by a uniform nearest-neighbor Hamiltonian, i.e.,
\begin{align}
H_{\text{L/R}}=&-\tau\sum_{n=1}^{\infty}
\sum_{ m,s}\left[d_{L/R,n+1,m,s}^{\dag}d_{L/R,n,m,s}\hspace{-0.6mm}+\text{H.c.}\right]~.
\end{align}
The eigenstates and energies of $H_\text{scatter}$ determine the Miller-Abrahams rates. Notice, however, that phonons are only present within the molecule. We thus restrict to spatial integral determining $\alpha_{i,j}$ accordingly. Moreover, immediately following a hop, the electron will not be in an eigenstate but populate only its projection inside the molecule. It may then either hop again after some time, or decay into the infinite lead. 

It thus remains to determine the Miller-Abraham rates as well as the life time. A standard way of imitating infinite leads is to replace them by finite ones with an imaginary energy $ \delta H = i \Gamma d_{L/R,n_\text{fin},m,s}^{\dag}d_{L/R,n_\text{fin},m,s}$ on the first and last sites~\cite{Datta}. This modification allows us numerically obtain all eigenstates and their decay rates $\nu^\text{decay}_{i,R/L}$ into the right or left lead.

The transmission probability through the molecule is determined by all possible paths that start with an electron injected via one of the leads. An electron in a given state $i$ has probabilities $\propto \nu^\text{decay}_{i,R/L}$ for decaying into the right or left lead where the path ends. Alternatively, it hops to the site $j$ with probability $\propto \nu_{i,j}$. Notice that the possibility of coherent transmission is included through paths with zero hops. We sample these trajectories using the Kinetic Monte-Carlo (KMC) method~\cite{bassler_1993} to obtain the transmission probabilities $T_{\sigma_\text{in},\sigma_\text{out}}(E_\text{in},E_\text{out})$. In the Supplementary material we give a detailed description of the modified KMC algorithm that we developed to include the infinite leads.

\begin{figure}
       \includegraphics[width=0.5\textwidth]{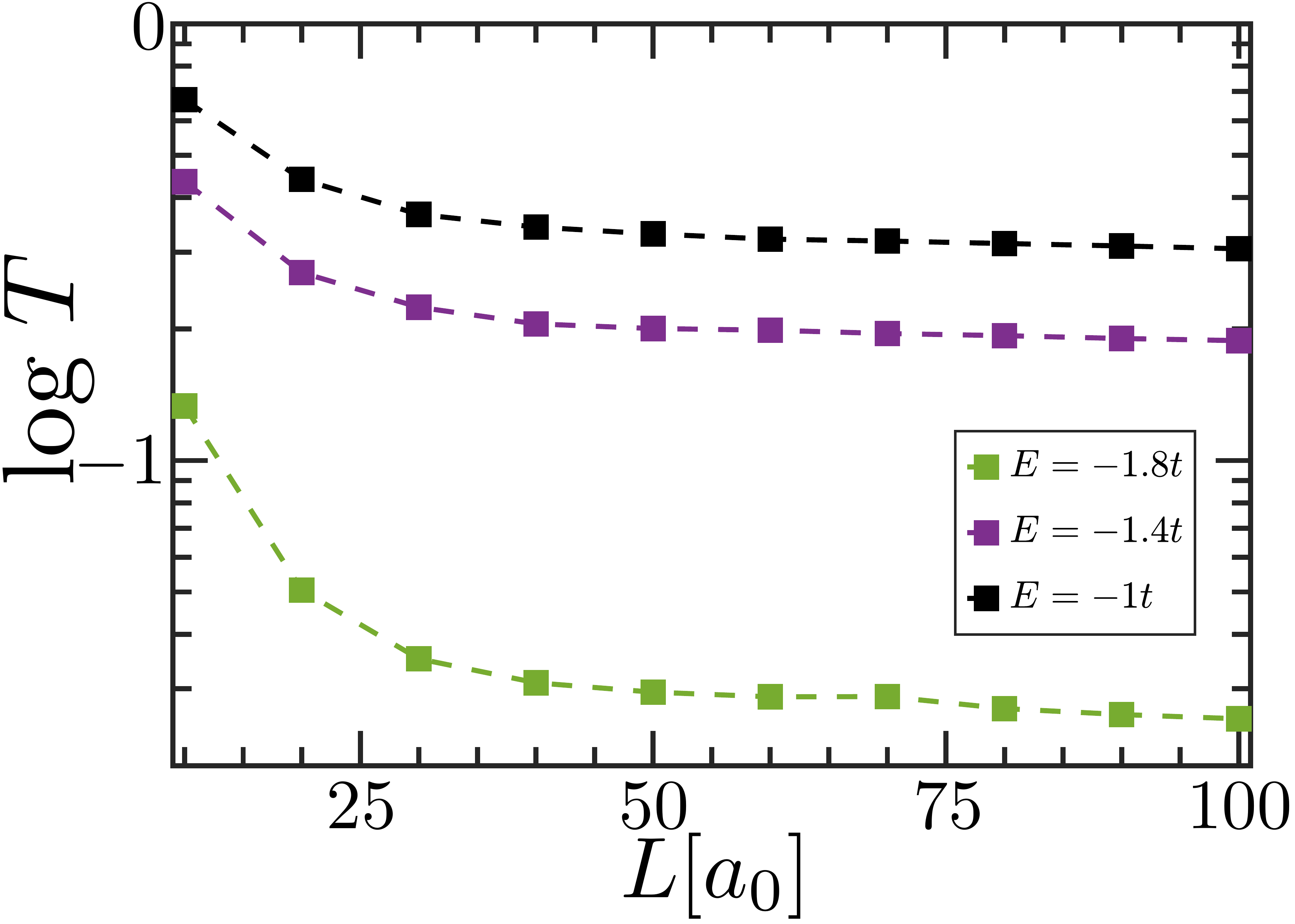}
                \caption{\textbf{Total hopping transmission in chiral molecules}. We show the disorder-averaged logarithm for incoming electrons with energy $E$ and arbitrary outgoing energy. For short molecules, the transmission decays exponentially, as for direct tunneling. For longer molecules it changes into a weaker length dependence that is consistent with a power law, as expected for phonon-assisted hopping.  }\label{fig6}
\end{figure}

Our approach is capable of describing the two main transport regimes and the crossover between them. In Fig.~\ref{fig6},  we present the total transmission as a function of the molecule's length for several values of $E$. At short distances, $T$ decreases exponentially with the length as expected in the tunneling dominant regime (see also the discussion in \textit{Localization in chiral structures}). At large distances, the transmission almost saturates and it is consistent with a weak power-law in $L$---a known signature of phonon-assisted hopping.

\subsection*{Spin-selective transmission in the hopping regime}

\begin{figure}
       \includegraphics[width=0.5\textwidth]{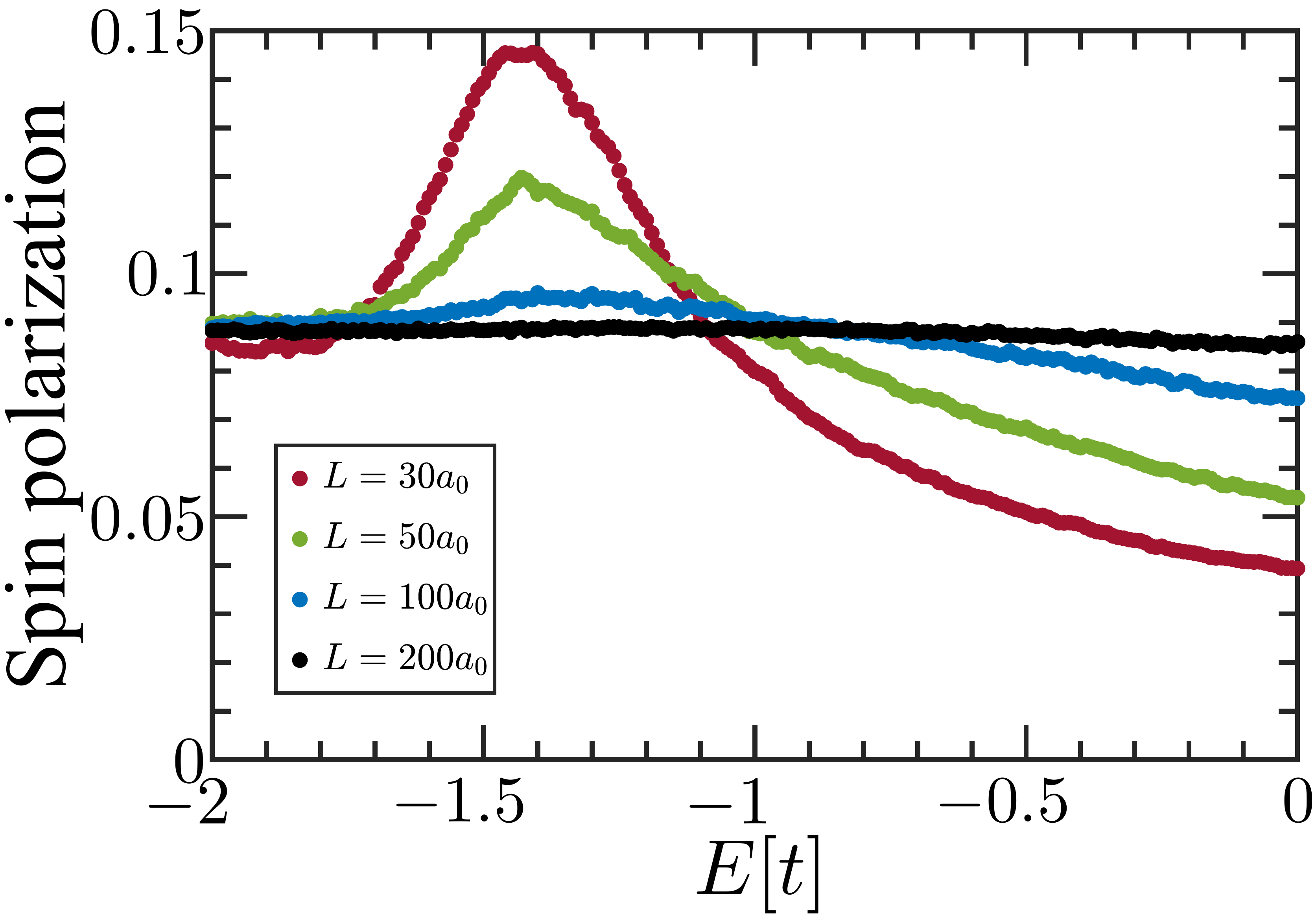}
                \caption{\textbf{Spin polarization in the VRH regime}. We present the spin polarization ${\cal P}(E)$ exhibited by electrons with initial energy $E$ and arbitrary spin after passing through molecules of different lengths. For short molecules, the data is similar to ${\cal P}^\text{coh}(E)$, see Fig.~\ref{fig3}. For longer molecules, ${\cal P}(E)$ saturates to an energy-independent value of nearly $10\%$. Importantly,  we found this substantial spin polarization without optimizing any parameters of the model.                  }\label{fig7}
\end{figure}

We study the spin-dependent transmission by computing the spin polarization of the outgoing electrons for an unpolarized injected beam, i.e.,
\begin{align}
{\cal P}(E)&\equiv\frac{\sum_{\sigma_\text{in},E_\text{out}}\left[T_{\sigma_\text{in},\uparrow}(E,E_\text{out})-T_{\sigma_\text{in},\downarrow}(E,E_\text{out})\right]}{\sum_{\sigma_\text{in},E_\text{out}}\left[T_{\sigma_\text{in},\uparrow}(E,E_\text{out})+T_{\sigma_\text{in},\downarrow}(E,E_\text{out})\right]}~.
\end{align}
Fig.~\ref{fig7} shows the energy dependence of ${\cal P}$ for different lengths. At short distances, the peak height and width are dictated by coherent tunneling and the height initially decreases with the length of the molecule. The width dramatically increases and results in a significant, energy-independent polarization for $L\gtrsim 200a_0$, where hopping is the primary transmission mechanism. This behavior arises from the combination of two ingredients: Energy-dependent spin selectivity near the boundary and energy lensing at the bulk of the molecule.

\begin{figure}
       \includegraphics[width=0.5\textwidth]{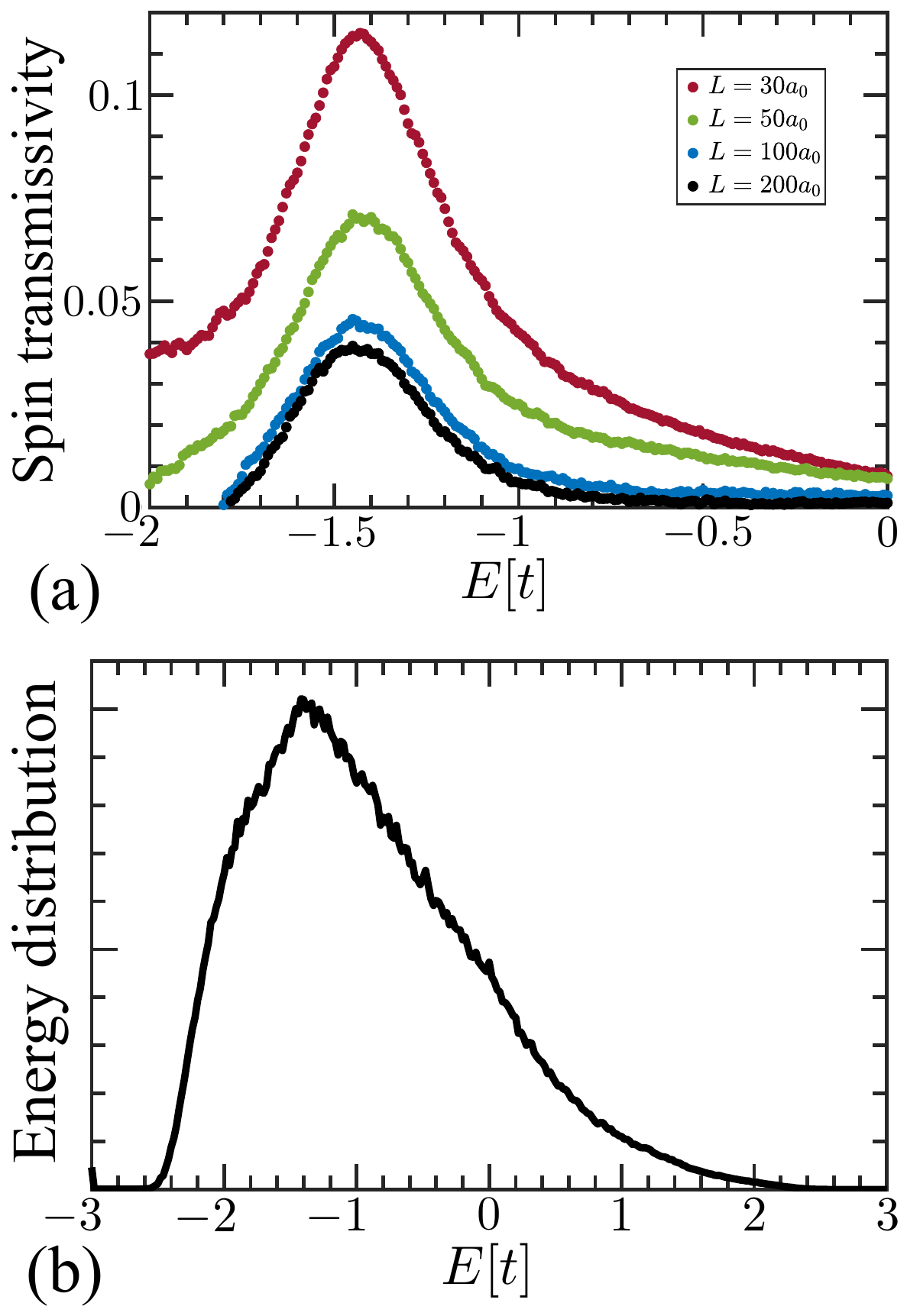}
                \caption{\textbf{Spin selectivity and energy lensing}. (a) We show  the spin transmissivity ${\cal S}(E)$ [see Eq.~\eqref{eqn.tranmissivity}] for molecules of different length. It saturates for sufficiently long molecules but, unlike ${\cal P}$, remains energy sensitive. 
               (b) We plot the energy distribution of outgoing electrons with arbitrary spin and evenly distributed initial energies after traversing a $L=200a_0$ molecule. Its peak defines the transport energy, which lies near the bottom of the band. These two quantities disentangle bulk and boundary contributions to ${\cal P}(E)$ as discussed in the text.    }\label{fig8}
\end{figure}

To disentangle the two effects, it is useful to look at a related quantity, the spin transmissivity
\begin{align}
{\cal S}(E)&\equiv\frac{\sum_{\sigma_\text{out},E_\text{out}}\left[T_{\uparrow,\sigma_\text{out}}(E,E_\text{out})-T_{\downarrow,\sigma_\text{out}}(E,E_\text{out})\right]}{\sum_{\sigma_\text{out},E_\text{out}}\left[T_{\uparrow,\sigma_\text{out}}(E,E_\text{out})+T_{\downarrow,\sigma_\text{out}}(E,E_\text{out})\right]}~.\label{eqn.tranmissivity}
\end{align}
It describes the total outgoing beam intensity for a spin-polarized incoming beam. The spin-transmissivity is independent from the polarization and thus carries separate information because phonon-assisted hopping is irreversible. We show its energy and length dependence in Fig.~\ref{fig8} (a). Similar to ${\cal P}$, it is determined by coherent tunneling for short molecules and eventually becomes length independent. However, in contrast to ${\cal P}$, the spin transmissivity remains narrowly peaked in energy (around the partial-gap energies) at all lengths. This behavior can be understood as follows: Without phonons, electrons with the preferred spin direction and in the appropriate energy window penetrate deeper into the molecule than those with opposite spin before being reflected. Consequently, their first hops typically occur over a longer distance. After the first few hops, the memory of the initial spin is entirely lost since localized states carry equal densities of either spin.


The same mechanism also polarizes the outgoing beam at the far end of the molecule. There, in addition, energy lensing is operative: after undergoing many hopping steps, the electron beam loses all memory of its initial energy. The \textit{outgoing} energies are instead peaked around the transport energy as shown in Fig.~\ref{fig8} (b). The large spin polarization independent of the \textit{incoming} beam energy thus arises from the overlap between the peaks of the spin-transmissivity [Fig.~\ref{fig8} (a)] and the transport energy [Fig.~\ref{fig8} (b)]. The former is determined by the periodicity of the helix-shaped molecule $\tilde{R}$. Modifying $\tilde{R}$ changes the  magnitude of the spin-polarization (see Fig.~S2 in the Supplementary material). Still, we expect that the two peaks are typically close to the bottom of the band, and hence, have a significant overlap.   

The lack of memory of both the incoming spin and energy  can explain several experimental findings: (i) The observation of CISS at applied voltages and/or temperatures much larger than the energy scale of the SOC. (ii) The weak dependence of the spin-polarization at the exit on the spin of the incoming current seen in Ref.~\cite{Goehler}. In the Supplementary material we show that similar ${\cal P},{\cal S}$ arise in time-of-flight calculations. Such quantities correspond to scattering experiments on molecules in the Coulomb blockade regime.

\subsection*{Magnetoresistance}

Above, we addressed the spin-polarization and transmissivity, which are relevant for CISS measurements by scattering through an empty band. Another standard method for observing spin-dependent transport is via current-voltage characteristics. In these experiments~\cite{Xie}, a voltage difference is applied accross the molecule via  one magnetic and one normal lead. We implement a magnetic lead by taking the imaginary on-site term $\delta H$ (and thus the density of states) spin dependent on that lead. Alternatively, one may include a magnetic barrier in the lead. We found similar results for the two methods and present data for the former.

In the presence of phonons, raising the chemical potential in one of the leads partially populates the states inside the molecule. The energy exchange between electrons and phonons establishes a local equilibrium distribution for each of these states. The charge transfer rate into any state $i$ (in the molecule or the leads) is thus given by~\cite{miller_1960} 
\begin{align}\label{CurrentVoltage}
I_{i}=\sum_{j}\left[\nu_{i,j}f_j(1-f_i)-\nu_{j,i}f_i(1-f_j)  \right],
\end{align}
where $f_{i}$ are the local distribution functions.
In the leads, they follow  the Fermi-Dirac distribution with chemical potentials $\mu_{\text{R}}$ and $\mu_{\text{L}}=\mu_{\text{R}}+eV$.
The occupations of the states inside the molecule are determined by the steady-state condition $I_{i}=0$ for all $i$ not in the leads. We numerically determine all local chemical potentials by solving this set of equations iteratively. Details of the solution are given in the Supplementary material.

The currents as a function of voltage for two opposite magnetizations are shown in Fig.~\ref{IV}. For the calculation we set $\Gamma=t$ for the majority spins as well as the non-magnetic lead, and   $\Gamma=t/2$ for the minority spins. We took the chemical potential on the right lead to be $\mu_{\text{R}}=-5t$,  well below the bottom of the conduction band in the clean limit. There is a significant difference between the two currents, which qualitatively resembles the experimental data~\cite{Xie}. The corresponding selectivity remains substantial as long as the chemical potential satisfies $\mu_{\text{L}}\lesssim-2t$. At larger chemical potential states near the middle of the band become occupied.  These states exhibit weaker spin selectivity but longer localization length and thus, dominate transport. Consequently, spin-polarization is supressed.

We emphasize that electron-phonon interactions give rise to a strongly non-linear dependence of the current on the applied voltage, which is essential for observing a non-zero magnetoresistance in a two-terminal setup~\cite{VanWees}.  Indeed, the selectivity sharply drops at small voltages. The precise functional dependence is difficult to ascertain numerically due to large relative errors in this regime.  
Our results for the spin-dependent scattering and magnetoresistance suggest that phonon-assisted hopping supports CISS significantly above the scale of SOC.

\subsection*{Conclusions}

We have studied CISS in the phonon-assisted hopping regime, pertinent for long molecules at room temperature. Our work shows that bulk energy lensing -- a well documented property of phonon assisted hopping -- plays a key role in enhancing the energy scale where CISS is operative. Specifically, we numerically computed the transmission probability in a scattering setup as well as the magnetoresistance in a transport configuration. Without optimizing any parameters of the model, we found substantial spin polarization on the order of 5-10\%. The mechanism described here is generic and does not rely on subtle details. It would be interesting to combine similar calculations with microscopic models of specific molecules.

More generally, we have demonstrated that CISS may be incoherent. In our model, full decoherence and spin relaxation occur on a scale $L_T$ much shorter than $L$, the length of the molecule. Since spin selectivity arises only from states that hybridize with the leads, the relevant scale for coherence is $\ell_\text{loc}$. We expect that full decoherence on even shorter scales, $L_T\ll \ell_\text{loc}$, will eventually destroy any spin dependence, as it does in the tunneling regime~\cite{Guo} already for $L_T \ll L$.

Finally, we note that our model does not include Coulomb repulsion between charge carriers. In semiconductors, such interactions are known to strongly modify VRH transport~\cite{ES}. The primary effect is due to the change in Coulomb potential induced by a single hop of a charge between two localized states. In addition, interactions can lead to Coulomb blockade in small systems, i.e., only a single carrier is permitted at a given time.  Scattering in this regime can be computed by a simple modification of our method: The various paths must be weighted by the time of flight (see Supplementary Material). We find that the resulting polarization is comparable and even slightly enhanced. A systematic study of interaction effects would be desirable in the future.

\subsection*{Acknowledgments} 
We thank Ron Naaman for many useful discussions and suggestions. This work was supported by Grant No. 2017608 from the United States-Israel Binational Science Foundation (BSF).

\end{document}


\begin{large}

\title{Incoherent chiral-induced spin selectivity - Supplementary Material}

\author{Lior Oppenheim}
\affiliation{Racah Institute of Physics, Hebrew University, Jerusalem 91904, Israel}
\affiliation{Department of Condensed Matter Physics, Weizmann Institute of Science, Rehovot 76100, Israel}
\author{Karen Michaeli}
\affiliation{Department of Condensed Matter Physics, Weizmann Institute of Science, Rehovot 76100, Israel}


\maketitle

\section{The kinetic Monte Carlo algorithm}\label{KMC}

This section describes the modified kinetic Monte Carlo (KMC) algorithm that we developed to calculate the transmission probability. Our method is based on the KMC algorithm introduced in Ref.~\cite{bassler_1993} for calculating particle diffusion in the phonon-assisted hopping regime. In the original method, particles are  injected into and collected from localized states. Consequently, the particles enter and leave the system with zero velocity. However, the CISS effect is highly sensitive to the electrons' velocity, or more precisely, their helicity. To capture it, the KMC simulation must start and end with extended states, where a velocity can be defined. Therefore, we study a system consisting of a molecule connected to two infinite leads. Such a modification to the algorithm is consistent with the measurement of CISS in scattering experiments. 

The eigenstates of a molecule connected to two infinite leads include (i) a finite number of localized states inside the molecule, and (ii) an infinite number of extended states in the leads. Wave functions of the latter describe scattering states that primarily reside in the leads, and hybridize with the molecule at the interface. To find both kinds of eigenstates numerically, we simulate the leads by two finite chains with an imaginary correction to the energy of the last site,  i.e., 
\begin{align}\label{Leads}\tag{S1}
\delta H = i \Gamma d_{L/R,n_\text{fin},m,s}^{\dag}d_{L/R,n_\text{fin},m,s}. 
\end{align}
Taking the leads' length to zero, which amounts to adding the imaginary energy to the first and last site of the molecule,  gives us the localized states alone. We denote these states by $\phi_{i,\sigma}(n)$, where $\sigma$ is the spin and $n$ the coordinate, and the index  $i = 1 \ldots 2N$. The corresponding eigenvalues are $\epsilon_{i}=E_{i}-i\eta_{i}$. The scattering states can be constructed from these eigenstates according to
 \begin{align}\label{Incoming}\tag{S2}
 \psi_{E,\sigma}(n)=\sum_{\sigma',n'}G_{E}(n,\sigma;n',\sigma')S(n',\sigma').
  \end{align}  
In contrast to $i$, the energy $E$ of the scattering states is continuous, and can be taken to be unbounded (for a lead with a non-zero density of states at all energies).  
 $G_{E}(n,\sigma;n',\sigma')$ is the Green's function describing the propagation of a particle with energy $E$ between point $n'$ and $n$, i.e.,
 \begin{align}\tag{S3}
G_{E}(n,\sigma;n',\sigma')=\sum_{i}\frac{\phi_{i,\sigma}(n)\bar{\phi}_{i,\sigma'}(n')}{E-E_{i}+i\eta_{i}}.
 \end{align}  
The source term $S(n,\sigma)=2i\Gamma\delta_{n,n_0}\delta_{\sigma,\sigma_{\text{in}}}$ describes a current injected at  the left ($n_0=1$) or right ($n_0=N$) end of the molecule. Such wave functions are commonly used for deriving scattering coefficients~\cite{Datta}.  In the case of coherent scattering, $R_{\sigma_{\text{in}},\sigma_{\text{out}}}^{\text{coh}}(E)=|\psi_{E,\sigma_{\text{out}}}(n_{\text{fin}}^L)-1|^2$ and $T_{\sigma_{\text{in}},\sigma_{\text{out}}}^{\text{coh}}(E)=|\psi_{E,\sigma_{\text{out}}}(n_{\text{fin}}^R)|^2$ are the reflection and transmission probabilities, and $\sum_{\sigma}\left[R_{\sigma_{\text{in}},\sigma_{\text{out}}}^{\text{coh}}(E)+T_{\sigma_{\text{in}},\sigma_{\text{out}}}^{\text{coh}}(E)\right]=1$. 

Implementing the KMC algorithm requires the total number of states available for hopping to be finite. Therefore, we had to reduce the number of scattering states. For this purpose, we took $n_{\text{fin}}=400$ and $n_{\text{fin}}=200$ for the source and the drain leads, respectively. To ensure that this length is sufficient, we verified that the results do not change as $n_{\text{fin}}^{L/R}$ is further increased. For our setup (with $n_{\text{fin}}^{L/R}>0$), the eigenstates $\phi_{i,\sigma}(n)$ include both localized and extended states. Although the energy of the extended states is no longer continuous, we still get a large local density of states near the interface between the molecule and the leads.

The hopping in each step of the simulation is determined by the Miller-Abrahams rates~\cite{miller_1960}:
\begin{align}\tag{S4}\label{MillerAbrahamsSM2}
	\nu_{i,j}=\nu_{0}\mathcal{A}_{i,j}\begin{cases} e^{-\beta\left(E_{j}-E_{i}\right)} & E_{j}>E_{i}\\1 & \text{else} \end{cases},
 \end{align}
 where
\begin{align}\tag{S5}
\label{OLSM2}
	\mathcal{A}_{i,j}=\sum_{n}\big{|}\sum_{\sigma}\bar{\phi}_{i}(n,\sigma)\phi_{j}(n,\sigma)\big{|}^2.
 \end{align}  
The sum over $n$ only includes sites that belong to the molecule and not those of the leads. These rates are used for determining the hopping at each step of the simulation. Moreover, they provide the typical time a particle spends in a state before it hops to the next one by absorbing/emitting phonons which is
\begin{align}\label{time}\tag{S6}
t_{i}=\frac{X}{\sum_{j}\nu_{j,i}},  
\end{align}  
where $X$ is a random (positive) Gaussian number with variance $1$. In the presence of the leads, we encounter an additional time scale: the typical time to escape from the system into one of the leads. This time scale is inversely proportional to the imaginary part of the eigenvalues, $\eta_{i}$. States that are strongly localized inside the system have negligible overlap with the leads and the imaginary part of their energies is likewise negligible. By contrast, the escape rates for states near the leads are significant. At each step of our modified KMC, we take into account both the Miller-Abrahams and escape probabilities to determine the next step of the simulation. The Miller-Abrahams  probabilities are given by
\begin{align}\tag{S7}
P_{i\rightarrow j}^{MA}=\frac{\nu_{j,i}}{\sum_{i}\nu_{i,j}}.
\end{align}
We estimate the escape probability as 
\begin{align}\tag{S8}
P_{i\rightarrow L/R}^{\text{exit}}=W_{L/R}(1-e^{-2\eta_{i} t_{i}}),
\end{align}
where
\begin{align}\tag{S9}
W_{L/R}=\frac{\sum_{\sigma}\sum_{n=1,n_{\text{fin}}^{L/R}}|\phi_{i}(n,\sigma)|^2
}{\sum_{\sigma}\left[\sum_{n=1,n_{\text{fin}}^L}|\phi_{i}(n,\sigma)|^2+\sum_{n=1,n_{\text{fin}}^R}|\phi_{i}(n,\sigma)|^2\right]}.
\end{align}

The discrete set of scattering states is sufficient for implementing the phonon-assisted hopping stage of the transport, and its termination where the particle escapes into the leads. For the initial step, we want to be able to (i) send a beam of particles at any energy $E$ into the system, and (ii) obtain the transmission probability both in the tunneling and in the hopping-dominated regimes. Therefore, we need to consider the infinite set of scattering states and we describe the initial state using  Eq.~\ref{Incoming}. The probabilities to hop from the initial state is determined by the Miller-Abrahams rates
\begin{align}\tag{S10}
\label{MillerAbrahamsSM}
	\nu_{i,E}=\nu_{0}\mathcal{A}_{i,E}\begin{cases} e^{-\beta\left(E-E_{i}\right)} & E>E_{i}\\1 & \text{else} \end{cases},
 \end{align}
 where
\begin{align}\tag{S11}
\label{OLSM}
	\mathcal{A}_{{i},E}=\sum_{n} \big{|}\sum_{\sigma}\bar{\phi}_{i}(n,\sigma)\psi_{E}(n,\sigma)\big{|}^2.
 \end{align} 
 Also here the sum over $n$ contains only the molecule's sites. 

Particles that hop from the injected beam lose coherence and can no longer contribute to the coherent reflection or transmission. Equivalently, the hopping can be viewed as leakage in the scattering problem with the rate $\sum_{j}\nu_{j,E}$. We calculate the probabilities to reflect $R_{\sigma_{\text{in}},\sigma_{\text{out}}}^{\text{coh}}(E)$, coherently transmit $T_{\sigma_{\text{in}},\sigma_{\text{out}}}^{\text{coh}}(E)$ to the opposite lead or hop by introducing   an imaginary energy $E\rightarrow E+ i\sum_{j}\nu_{j,E}$ into the calculation of  $\psi_{E}(n,\sigma)$ in Eq.~\ref{Incoming}. Then, for particles entering the molecule from the left lead, the probabilities to reflect, tunnel and hop are
\begin{subequations}
\begin{align}\label{R}\tag{S12a}
\hspace{-4mm}R_{\sigma_{\text{in}},\sigma_{\text{out}}}^{\text{coh}}(E)=|\psi_{E+ i\sum_{j}\nu_{j,E}}(n_{L,\text{fin}})-1|^2;
\end{align}
\begin{align}\label{T}\tag{S12b}
\hspace{-27mm}T_{\sigma_{\text{in}},\sigma_{\text{out}}}^{\text{coh}}(E)=|\psi_{E}(n_{R,\text{fin}})|^2;
\end{align} 
\begin{align}\label{P}\tag{S12c}
P_{\sigma_{\text{in}}}^{hop}(E)=1-\sum_{\sigma}\left[R_{\sigma_{\text{in}},\sigma}^{\text{coh}}(E)+T_{\sigma_{\text{in}},\sigma}^{\text{coh}}(E)\right].
\end{align} 
\end{subequations}
Notice that $\sum_{\sigma}\left[R_{\sigma_{\text{in}},\sigma_{\text{out}}}^{\text{coh}}(E)+T_{\sigma_{\text{in}},\sigma_{\text{out}}}^{\text{coh}}(E)\right]\neq1$, i.e., the imaginary energy results in breakdown of charge conservation (unitarity) in the scattering problem. The missing particles are those that enter the hopping stage.
\\
\\
\textbf{The kinetic Monte Carlo algorithm}
The aim of the algorithm is to determine the probability of electrons with energy $E_{\text{in}}$ and spin $\sigma_{\text{in}}$ to be transmitted between the leads. We collect information about the spin and energy at the final step, and about the time-of-flight. The simulation is repeated for different realizations of the disorder  potential (at least $10^{4}$ times). To improve convergence, we average the logarithm of the probability over different realizations. For each energy, spin, and realization we perform the following steps. Without loss of generality, we consider here particles coming from the left lead and calculate their probability to exit through the right lead. 

 \begin{enumerate}
 \item[\textbf{1.}] \textbf{Diagonalization}:\\ Diagonalize the Hamiltonian $H_{\text{scatter}}=S^{-1}DS$ (Eq.~8 in the main text). The eigenstates $\phi_{i}(n,\sigma)$ are the raws of the diagonalizing matrix $S$ and $\bar{\phi}_{i}(n,\sigma)$ are the raws of $S^{-1}$.

\item[\textbf{2.}] \textbf{Injection stage}:\\ 
Calculate the Miller-Abrahams hopping rates $\nu_{i,E}$ from the initial scattering state to all eigenstates using Eq.~\ref{MillerAbrahamsSM}. Randomly select the next state according to the probabilities $P_{E\rightarrow i}=\frac{\nu_{i,E}}{\sum_{i}\nu_{i,E}}$. For the time-of-flight calculation, record the transition time $t_{E}=\frac{X}{\sum_{i}\nu_{i,E}}$, where $X$ is a random (positive) Gaussian number with variance $1$.

 \item[\textbf{3.}] \textbf{Probability to enter the hopping regime}:\\ Compute the scattering wave function (Eq.~\ref{Incoming}) at the complex energy, $\psi_{E_{\text{in}}+ i\sum_{j}\nu_{j,E},\sigma_{\text{in}}}(n)$. Use the wave function to determine the reflection (Eq.~\ref{R}), transmission (Eq.~\ref{T}) and hopping (Eq.~\ref{P}) probabilities.

\item[\textbf{4.}] \textbf{Hopping rates in the bulk}:\\ Starting at state $j$, compute the Miller-Abrahams probabilities $P_{i\rightarrow j}=\frac{\nu_{j,i}}{\sum_{j}\nu_{j,i}}$, where $\nu_{j,i}$ is given by Eq.~\ref{MillerAbrahamsSM2}. 

 \item[\textbf{5.}] \textbf{Determine the next state}:\\ 
With probability $1-P_{\text{exit},L}-P_{\text{exit},R}$ the particle continues hopping and the simulation proceeds at step 4. [For the time-of-flight calculation, record the transition time $t_{i}$ defined in Eq.~\ref{time}.] Otherwise, the simulation ends. The typical time for the last step is $1/\eta_j$. Store the density of the spin up and down component of the particle at the right lead, $\rho_{\sigma_{\text{out}}}(E_j)=\frac{\sum_{n=1,n_{\text{fin}}^R}|\phi_{j}(n,\sigma)|^2}{\sum_{n,\sigma}|\phi_{j}(n,\sigma)|^2}$, as well as the energy at the last stage $E_j$.

 \end{enumerate}

For each incoming energy $E$ and spin $\sigma_{\text{in}}$, we repeated the simulation $10^{4}$ times for every realization of the disorder potential and for $6000$ different realizations. Then, we calculated the spin-dependent transmission probability per simulation is 
\begin{align}\tag{S13}
T_{\sigma_{\text{in}},\sigma_{\text{out}}}(E_{\text{in}},E_{\text{out}})=T_{\sigma_{\text{in}},\sigma_{\text{out}}}^{\text{coh}}(E_{\text{in}},E_{\text{out}})+P_{\sigma_{\text{in}}}^{hop}(E_{\text{in}})\rho_{\sigma_{\text{out}}}(E_{\text{out}}).
\end{align}

\section{The sensitivity of the spin-polarization to the helix periodicity}

In the main text, we discuss the combined effect of energy lensing in the bulk and spin selectivity at the boundaries. An important aspect is that efficient spin-polarization requires a strong boundary spin effect at the transport energy. Since both the transport energy and the peak in spin-transmissivity occur close to the bottom of the band, we expect significant spin-polarization universally for all chiral molecules. In Fig.~\ref{fig2SM} we show that the spin-selectivity remains significant as the periodicity of the helix, $\tilde{R}$, changes by a factor of more than two. A decrease in the CISS effect is, however, expected as $\tilde{R}$ grows and the molecule loses its chiral properties. Indeed, we find  that both the transmissivity and the polarization become smaller for $\tilde{R}=8$. 

\begin{figure}\center{
       \includegraphics[width=0.5\textwidth]{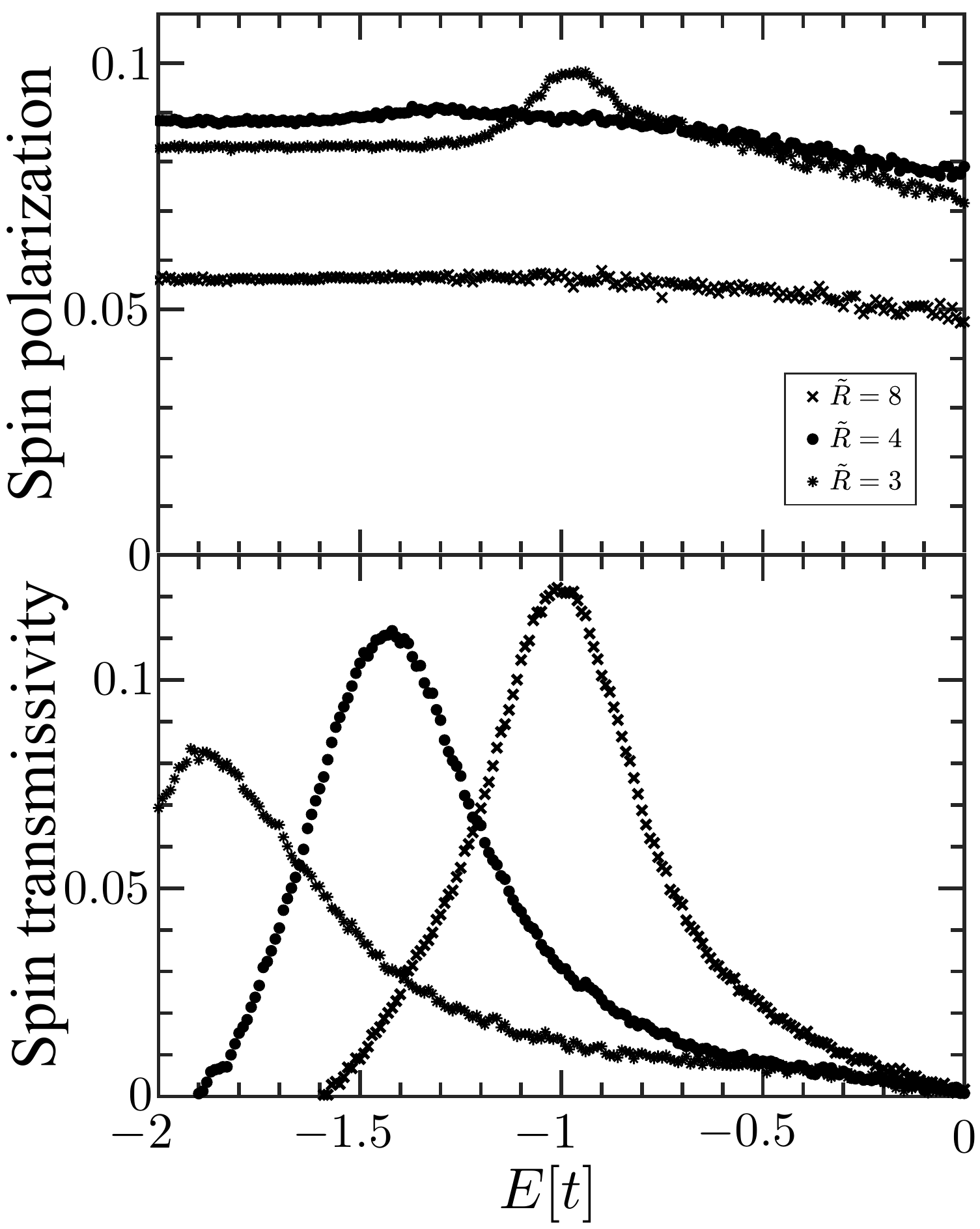}
                \figuretag{S1}\caption{\textbf{CISS in the hopping regime for molecules of different curvature}. We present the spin polarization ${\cal P}(E)$ and transmissivity ${\cal S}(E)$ 
                for molecules of length $L=100$ with different helical periodicity $\tilde{R}$. The spin dependence moderately decreases as $\tilde{R}$ grows, but remains significant as long as the curvature is not too small.                
                  }\label{fig2SM}}
\end{figure}

\section{Time of flight calculation}

\begin{figure}\center{
       \includegraphics[width=0.5\textwidth]{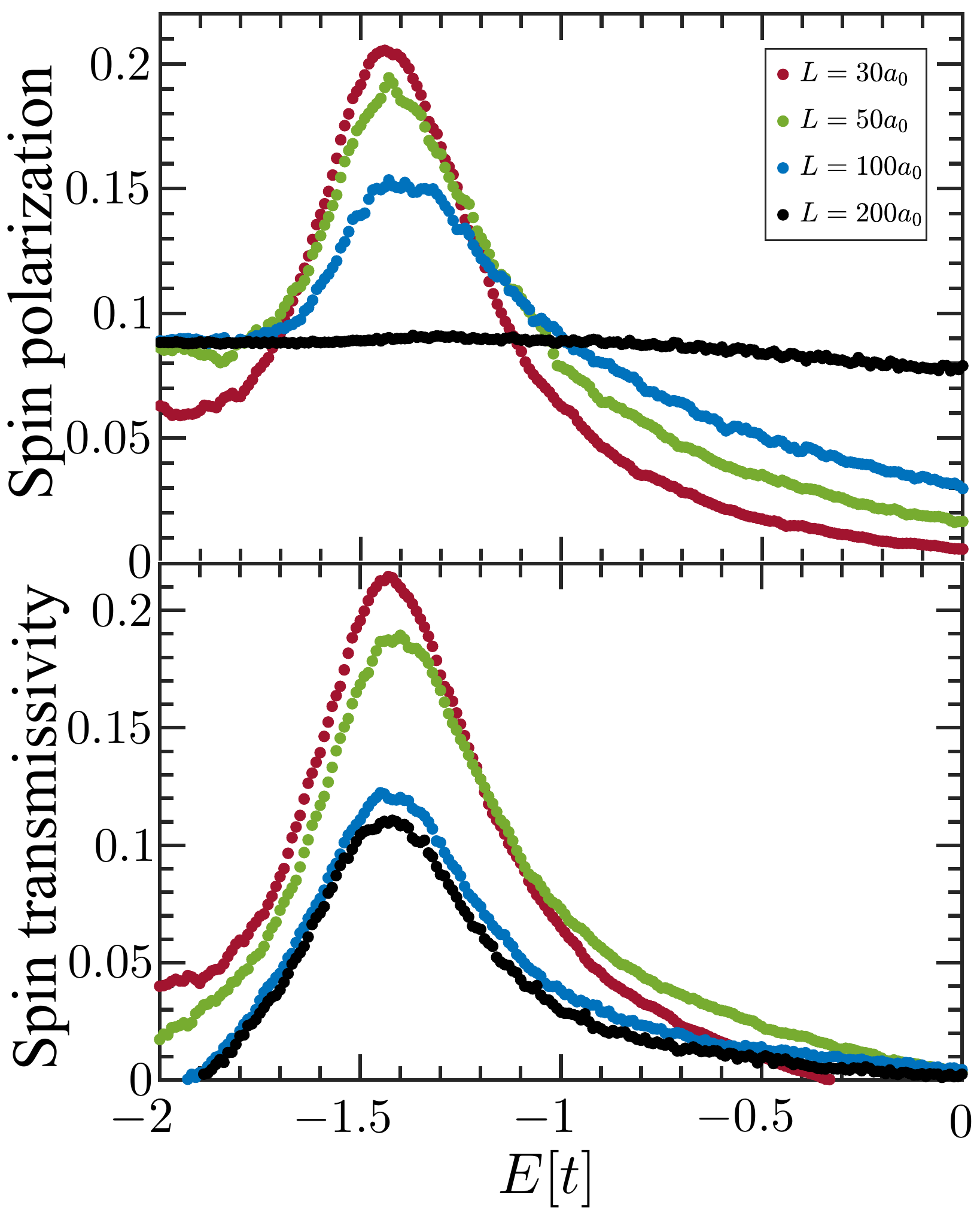}
               \figuretag{S2}  \caption{\textbf{CISS in the electron time-of-flight}. We present the spin polarization ${\cal P}(E)$ and transmissivity ${\cal S}(E)$ 
                calculated from the time-of-flight of electrons through a chiral molecule of length $L=100$. While for short molecules, the two quantities are larger than those found from the transmission, they become comparable at longer lengths  (see Figs.~5 and 6).               
                  }\label{fig3SM}}
\end{figure}

Above, we described our modified KMC algorithm. We presented the steps in the calculation of the transmission probability.  The first is directly related to the scattering experiments where the only important interaction is the one between electron and phonons that facilitates hopping. If interactions between electrons and localized charges inside the molecule (electrons or nuclei) are strong, their effect on the hopping processes must be included. In particular, if the interaction prevents  two electrons from entering the molecule at any given time, i.e., the system is in the Coulomb blockade regime, the scattering experiment no longer measures only the transmission probability. Instead, the current intensity exiting the molecule is modified by the time each electron spends passing through the molecule, i.e., by its time-of-flight.  While computing the KMC algorithm, we keep the typical time between hops at each step (see description in the previous section). At the end of the simulation we can calculate the time-of-flight
\begin{align}\tag{S14}
TOF_{\sigma_{\text{in}},\sigma_{\text{out}}}(E_{\text{in}},E_{\text{out}})=\tau T_{\sigma_{\text{in}},\sigma_{\text{out}}}^{\text{coh}}(E_{\text{in}},E_{\text{out}})+P_{\sigma_{\text{in}}}^{hop}(E_{\text{in}})\rho_{\sigma_{\text{out}}}(E_{\text{out}})\sum_j t_{j}^{-1},
\end{align} 
and average it over different runs of the simulation. Here, the final terms sums over the entire path taken by the particle between the two leads.

In Fig.~\ref{fig3SM},  we show the spin-polarization and transmissivity of the electronic time-of-flight. For long molecules, we see no difference between the spin-polarization obtained from the time-of-flight and transmission calculations. The time-of-flight gives rise to enhanced spin-transmissivity (and polarization of short molecules), suggesting that interactions can further increase CISS also in the phonon-assisted hopping regime.

\section{Calculation of the current-voltage characteristics}

We calculate the current-voltage characteristics in a two-terminal setup, i.e., the molecule is connected to two leads at equilibrium with chemical potential $\mu_{L/R}$.  The occupancy of the scattering state with energy $E$ inside the left (right) lead is given by the Fermi-Dirac distribution $f_{L/R}(E)=[e^{\beta(E-\mu_{L/R})}+1]^{-1}$.  As a consequence of energy exchange between the electrons and phonons, each electronic state $j$ inside the molecule establishes local equilibrium, and it acquires an average occupancy $f_{j}$. The current into a state $i$ is then given by Eq.~12 ($i$ can be a state inside the molecule or the leads).

The measurement of the current is performed at the leads:
\begin{align}\label{LeadsCurrent}\tag{S15}
I_{L/R}&=\int\frac{dE}{2\pi}\left[\nu_{E,i}f_{i}(1-f_{L/R}(E))-\nu_{i,E}(1-f_{i})f_{L/R}(E)\right]\\\nonumber
&+\int\frac{dEdE'}{4\pi^2}\left[\nu_{E,E'}f_{R/L}(E')(1-f_{L/R}(E))-\nu_{E',E}(1-f_{R/L}(E'))f_{L/R}(E)\right].
\end{align}
Here, $\nu_{E,i}$ are the Miller-Abrahams rates given in Eq.~\ref{MillerAbrahamsSM} and $\nu_{E,E'}$ is the phonon-assisted hopping rate between the leads
\begin{align}\label{MillerAbrahamsLeadLead}\tag{S16}
	\nu_{E',E}=\nu_{0}\mathcal{A}_{E',E}\begin{cases} e^{-\beta\left(E-E'\right)} & E>E'\\1 & \text{else} \end{cases},
 \end{align}
 where
\begin{align}\tag{S17}
\label{OLSMLL}
	\mathcal{A}_{E',E}=\sum_{n} \big{|}\sum_{\sigma}\bar{\psi}_{E'}(n,\sigma)\psi_{E}(n,\sigma)\big{|}^2.
 \end{align} 
Notice  that we assume that the molecule is long enough that the direct tunneling between the leads is negligible. 

To determine the current, we need the average occupancies $f_j$ of the states inside the leads. eigenstates of $H_{scatter}$. These quantities can be found from the steady state condition~\cite{miller_1960} that no current accumulates in these states ($I_{i\in \text{mol}}=0$), i.e.,
\begin{align}\label{BulkCurrent}\tag{S18}
\int\hspace{-.5mm}\frac{dE}{2\pi}\hspace{-.5mm}\left[\nu_{i,E}(1-f_{i})f_{L/R}(E)-\nu_{E,i}f_{i}(1-f_{L/R}(E))\right]\hspace{-.5mm}+\hspace{-.5mm}\sum_{j\neq i}\left[\nu_{i,j}(1-f_{i})f_{j}-\nu_{j,i}f_{i}(1-f_{j})\right]\hspace{-.5mm}=\hspace{-.5mm}0
\end{align}
All Miller-Abrahams rates into and out of the leads enter the equations under two integrals over the energy $E$ weighted with $f_{L/R}(E)$ and $1-f_{L/R}(E)$. As a result, we can take the length of the leads to zero (see discussion in Section~\ref{KMC}), compute the infinite set of extended states inside the leads given by Eq.~\ref{Incoming}, and perform the integrals. The integrands include sums over resonances of various widths, and therefore the numerical integral must be performed carefully. The magnetoresistance experiments are performed by connecting the molecule to one magnetic lead. Then the current as a function of voltage is measured when the magnet is once aligned parallel and once anti-parallel to the molecular axis. To describe such a lead, we change the imaginary correction to the Hamiltonian (Eq.~\ref{Leads}) to be spin- and lead-dependent. Then, we take $\Gamma_{\sigma}^{R}=t$, $\Gamma_{\sigma_0}^{L}=t$, and  $\Gamma_{-\sigma_0}^{L}=t/2$, where $\sigma_0$ is in the spin projected into the direction of the magnet. 

We solve Eq.~\ref{BulkCurrent} for the occupancies inside the molecule iteratively, i.e., we compute $f_{i}^{(n+1)}=f_{i}^{(n)}+\delta f_{i}$, with $f_{i}^{(0)}=0$. We stop the iterations when $|(I_{R}+I_{L})/(I_{R}-I_{L})|<0.01$, with $I_R$ and $I_L$ determined by Eq.~\ref{LeadsCurrent}. This condition reflects the fact that charge is conserved and all the current leaving the left lead must reach the right one.

\end{large}